\documentclass[aps,prb,preprint,showkeys,floatfix]{revtex4}

\usepackage{graphicx,color}
\usepackage{microtype}
\usepackage{multirow,slashbox}
\usepackage{natbib}
\usepackage[utf8]{inputenc}
\usepackage{float}
\graphicspath{{figs/}}
\usepackage{amsmath}
\usepackage{tabularx}
\usepackage[colorlinks,linkcolor=black,anchorcolor=black,citecolor=black]{hyperref}
\usepackage{amssymb}
\usepackage{subfigure}
\date{\today}

\begin{document}

\title{Buckling of graphene/MoS$_2$ van der Waals heterostructures: the misfit strain effect}

\author{Run-Sen Zhang}
    \affiliation{Shanghai Key Laboratory of Mechanics in Energy Engineering, Shanghai Institute of Applied Mathematics and Mechanics, School of Mechanics and Engineering Science, Shanghai University, Shanghai 200072, People's Republic of China}

\author{Jin-Wu Jiang}
    \altaffiliation{Corresponding author: jwjiang5918@hotmail.com}
    \affiliation{Shanghai Key Laboratory of Mechanics in Energy Engineering, Shanghai Institute of Applied Mathematics and Mechanics, School of Mechanics and Engineering Science, Shanghai University, Shanghai 200072, People's Republic of China}

\date{\today}
\begin{abstract}

Van der Waals heterostructures are constructed by stacking different atomic layers and can inherit many novel electronic and optical properties from the constituting atomic layers. Mechanical stability is of key importance for the high performance of nano devices based on the van der Waals heterostructure. In particular, buckling instability is a critical mechanical issue for the heterostructure due to its two-dimensional nature. Using graphene/MoS$_2$ heterostructure as an example, the present work demonstrates the relationship between the buckling instability and the inevitable misfit strain in the heterostructure by molecular dynamics simulations. The misfit strain has rather different effects on the buckling phenomenon depending on the magnitude of the misfit strain. (1). For negative misfit strain, the buckling stability of the heterostructure is reduced by the misfit strain. It is because the graphene layer, which initiates the buckling process in the heterostructure of negative misfit strain, is pre-compressed by the misfit strain that accelerates the buckling of the graphene layer. (2). For small positive misfit strain, the buckling stability for the graphene/MoS$_2$ heterostructure is elevated. The underlying mechanism is that the graphene layer initiates the buckling process of the heterostructure and is pre-stretched by the small positive misfit strain, which decelerates the buckling of the graphene layer. (3). For large positive misfit strain, the graphene layer is pre-stretched while the MoS$_2$ layer is considerably pre-compressed, so the buckling of the heterostructure is initiated by the MoS$_2$ layer. As a consequence, the buckling stability of the graphene/MoS$_2$ heterostructure is reduced by the increase of large positive misfit strain. These findings shall be valuable for understanding mechanical properties of van der Waals heterostructures.

\end{abstract}
\keywords{van der Waals heterostructure, misfit strain, buckling}

\maketitle

\section{Introduction}


Two-dimensional (2D) van der Waals heterostructures have attracted great scientific research interest in recent years due to their outstanding functional properties inherited from different constitute layers.\cite{Novoselov20162D,Liu2016Van} Compared with bulk materials, 2D materials have ultra low bending rigidity and can be easily buckled or wrinkled under compressive load in the in-plane direction. Taking advantage of the buckling or wrinkling properties, 2D materials are promising candidates for stretchable and flexible electronics, functional surface bionics, soft sensors, actuators, and etc\cite{Ko2020All,2016Reversible}.

The buckling of single atomic layers like graphene or MoS$_2$ have been extensively studied. It has been illustrated that buckling properties of graphene are sensitive to various effects, including the initial ripple,\cite{Xiang2016Compressive} the defect,\cite{Tserpes2015Buckling} the size effect and biaxial compression,\cite{Pradhan2010Small,Shi2011Nonlocal} and the shear loading.\cite{Wen2011Controlling,Zhang2012Tunable,Zhou2015Mechanics} Jiang investigated the effect of strain rate and temperature on the critical buckling strain of the single layer MoS$_2$ sheet under uniaxial compression.\cite{Jiang2014The} The fan-blade shaped wrinkling of single layer MoS$_2$ under circular torsion was studied by Bao et al..\cite{Bao2017Circular} The intrinsic thermal vibration induced ripples are closely related to the stiffness and bending modulus for the MoS$_2$.\cite{Singh2015Rippling} The defect can reduce the buckling stability of MoS$_2$ under uniaxial compressive loadings.\cite{Li2019Molecular}

As a characteristic feature of the van der Waals heterostructure, misfit strain between the constituting atomic layers is inevitable, due to different lattice constants of the atomic layers.\cite{Jiang2014Mechanical,Lin2014Atomically,Linyang2014Structures} The misfit strain in the heterostructure can be controlled through various methods, including the substrate,\cite{Ni2008Raman} the structural design,\cite{Chaste2018Nanomechanical} and the photoelectrochemical etching process.\cite{shivaraman2013Raman} The misfit strain can cause direct effects on physical and mechanical properties for the van der Waals heterostructure. For instance, it was found that the misfit strain induces additional energy dissipation for the mechanical resonant oscillation of the graphene/MoS$_2$ heterostructure.\cite{He2019Misfit}

Although the buckling behavior of individual atomic layers has been extensively studied, the buckling phenomenon for the van der Waals heterostructures is not well studied. In particular, the effect of the inevitable misfit strain on the buckling of van der Waals heterostructure is still unclear. The mechanical stability, especially the buckling stability, is a key factor for stable performance of nano devices based on the van der Waals heterostructure. It is thus an urgent task to investigate the mechanism for the buckling of heterostructure with different misfit strains.

In this paper, we perform molecular dynamics (MD) simulations to investigate the buckling behavior of graphene/MoS$_2$ van der Waals heterostructures subjected to compressive loadings. We find that the misfit strain within the graphene/MoS$_2$ heterostructure has significant effect on the buckling phenomenon. There are three typical ranges for the misfit strain. For negative misfit strain, the graphene layer in the heterostructure is pre-compressed, so the graphene layer initiates the buckling process of the heterostructure. Hence, the critical buckling strain for the heterostructure is considerably reduced by the negative misfit strain. For small positive misfit strain, the buckling of the heterostructure is still initiated by the graphene layer which is slightly pre-stretched by the misfit strain, so the critical buckling strain for the heterostructure is enhanced by small positive misfit strain. For large positive misfit strain, the MoS$_2$ layer is pre-compressed seriously and it initiates the buckling process of the heterostructure, so the critical buckling strain will be reduced by increasing misfit strain in this range. We also show that the buckling direction for the heterostructure is also affected by the misfit strain.

\section{Structure and simulation details}

\begin{figure}[htbp]  
  \begin{center}
    \scalebox{1.0}[1.0]{\includegraphics[scale=0.3]{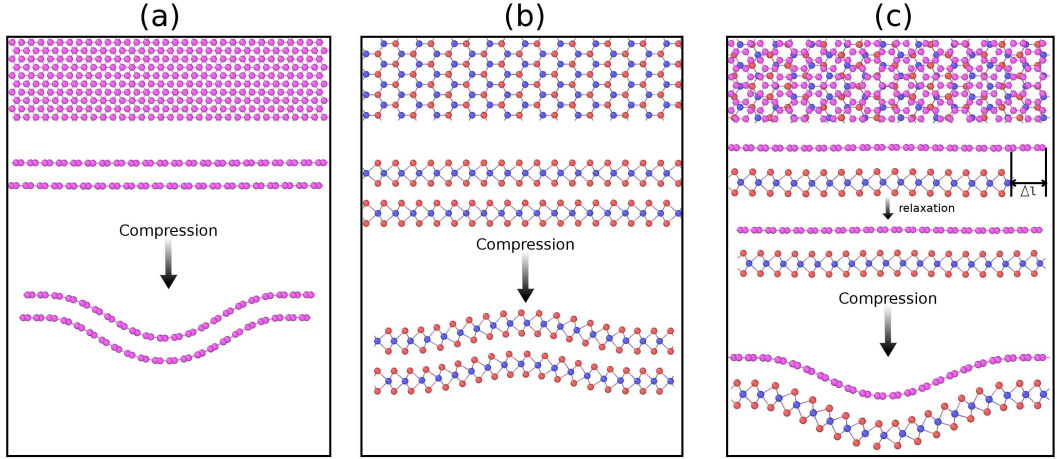}}
  \end{center}
  \caption{Bilayer structures studied in the present work. (a) The graphene/graphene structure is compressed and buckled. (b) The MoS$_2$/MoS$_2$ structure is compressed and buckled. (c) The graphene/MoS$_2$ heterostructure. The graphene and MoS$_2$ have a difference of $\Delta l$ in their length, resulting in a misfit strain in the relaxed configuration. The heterostructure is compressed and buckled.}
  \label{fig_model}
\end{figure}

In the present work, three structures shown in Fig.\ref{fig_model} are comparatively investigated. The graphene/graphene bilayer structure has the AB stacking order. The MoS$_2$/MoS$_2$ bilayer structure is in its lowest-energy stacking order. These structures are compressed by uniaxial compressive strains along the length direction, and will be buckled when the strain is above a critical value. The buckling property is sensitive to the length of the structure, while the width is not important for the buckling phenomenon. We thus choose a constant width for graphene and MoS$_2$ for all structures simulated in the present work. The width of graphene is 12.3~{\AA} and the width of MoS$_2$ is 12.5~{\AA}. These widths are properly chosen, so that the misfit strain in the width direction is minimum. The length is along the armchair direction, and the width is along the zigzag direction.

All MD simulations are performed by the Large-scale Atomic/Molecular Massively Parallel Simulator (LAMMPS) code,\cite{plimpton1995fast} and the OVITO package is employed for visualization.\cite{stukowski2009visualization} The standard Newton equations of motion are integrated in time using the velocity Verlet algorithm with a time step of 1~{fs}. All simulations are performed at the low temperature of 4.2~{K}, so that the effect of thermal vibration is less important. Periodic boundary conditions are applied in the two in-plane directions, while the free boundary condition is applied in the out-of-plane direction. The structures are uniaxially compressed along horizontal length direction with a strain rate of $10^{7}$~{s$^{-1}$}, which is a typical value used in many previous works.\cite{Jiang2014The,Tserpes2015Buckling,Xiang2016Compressive}

The carbon-carbon interactions are described by the second generation Brenner potential,\cite{Brenner2002A} while the MoS$_2$ interatomic interactions are described by the Stillinger-Weber potential.\cite{Jiang2013Molecular} The inter-layer interactions between graphene/graphene, MoS$_2$/MoS$_2$ and graphene/MoS$_2$ are described by the Lennard-Jones potential. The distance and energy parameters in the Lennard-Jones potential are listed in table~\ref{tab_lj}.
\begin{table}[htbp]
\caption{Lennard-Jones parameters used in the present work for graphene/MoS$_2$,\cite{Jiang2014Mechanical} graphene/graphene,\cite{Jiang2015A} and MoS$_2$/MoS$_2$.\cite{Jiang2015A}}
\label{tab_lj}
\begin{tabular}{@{\extracolsep{\fill}}|c|c|c|c|}
\hline 
  structure  & $\epsilon$~{(meV)} & $\sigma$~({\AA}) & cut off~({\AA})\tabularnewline
\hline 
\hline 
graphene/MoS$_2$ & 3.95 & 3.625 & 10.0\tabularnewline
\hline 
graphene/graphene & 2.96 & 3.382 & 10.0\tabularnewline
\hline
MoS$_2$/MoS$_2$ & 23.6 & 3.18 & 10.0\tabularnewline
\hline
\end{tabular}
\end{table}

Our simulations are performed as follows. The system is first thermalized for 1~{ns} within the NPT (i.e. the particles number N, the pressure P and the temperature T of the system are constant) ensemble by the Nose-Hoover thermostat.\cite{nose1984unified,hoover1985canonical} Then the structures are compressed uniaxially in the horizontal length direction, while the lateral directions are allowed to be fully relaxed during the compression process. A value for the thickness is required in the computation of the stress. The thickness is not well defined for atomic thick materials like graphene and MoS$_2$, so we have used the same thickness of 10.0~{\AA} for all of these three types of structures studied in the present work.

\section{Results and discussion}

\begin{figure}[b]  
  \begin{center}
    \scalebox{1.0}[1.0]{\includegraphics[scale=1.0]{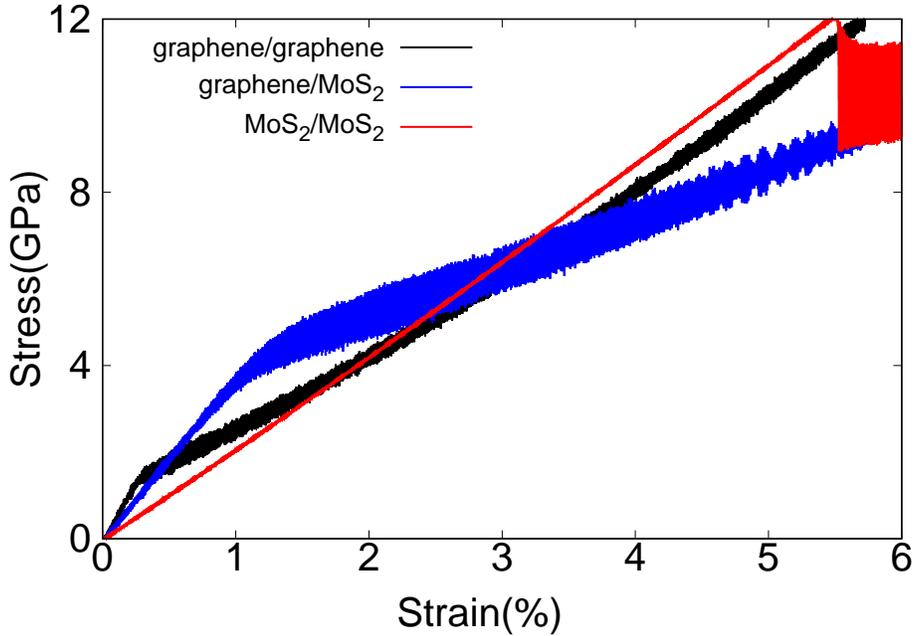}}
  \end{center}
  \caption{The stress-strain relations for graphene/graphene, MoS$_2$/MoS$_2$, and graphene/MoS$_2$ heterostructures under uniaxial compression.}
  \label{fig_2}
\end{figure}

\begin{figure}[htpb]  
  \begin{center}
    \scalebox{1.0}[1.0]{\includegraphics[scale=1.0]{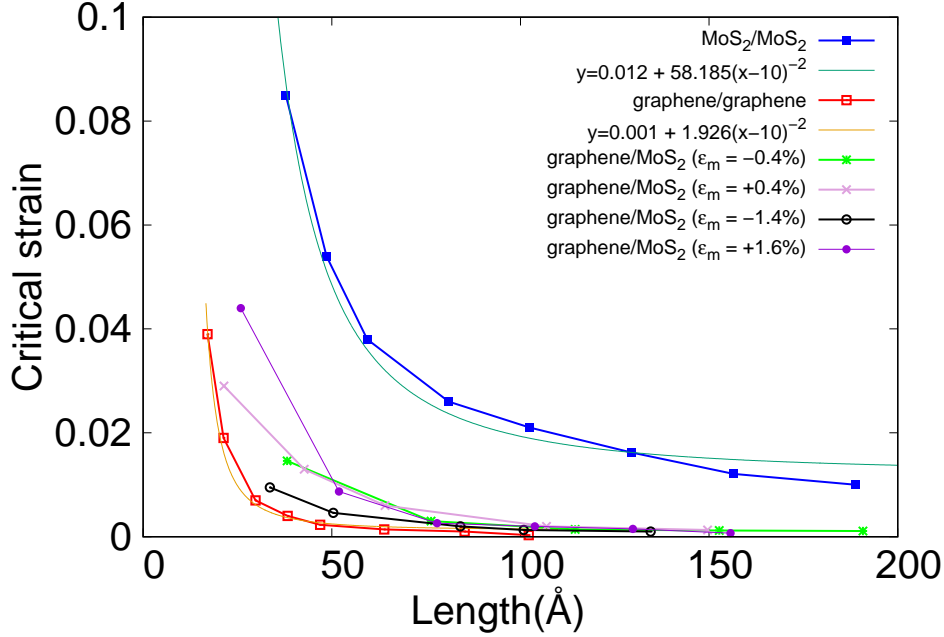}}
  \end{center}
  \caption{The length dependence for the critical buckling strain of graphene/graphene, MoS$_2$/MoS$_2$, and the graphene/MoS$_2$ heterostructure with different misfit strain $\epsilon_m$.}
  \label{fig_lsi}
\end{figure}

Figure~\ref{fig_2} shows the stress-strain relation for these three structures with similar length around 47~{\AA}. The buckling strains are 0.22\%, 5.5\%, and 1.1\% for the graphene/graphene bilayer, MoS$_2$/MoS$_2$ bilayer, and the graphene/MoS$_2$ heterostructure, respectively. The buckling strain for the MoS$_2$/MoS$_2$ structure is considerably larger than the graphene/graphene structure. This large difference can be understood by the Euler buckling theory. According to Euler buckling theory, the critical buckling strain of a thin plate is\cite{TimoshenkoS1987},
\begin{eqnarray}
\epsilon_c=-\frac{4\pi^2D}{EL^2},
\end{eqnarray}
where $D$ is the bending modulus, $E$ is the in-plane stiffness, and $L$ is the length. By linear fitting of the stress-strain relation within [0, 0.2\%] in Fig.~\ref{fig_2}, we get the stiffness of 552~{GPa} and 210~{GPa} for the graphene/graphene and MoS$_2$/MoS$_2$, respectively. The bending modulus for a single MoS$_2$ layer is larger than a single graphene layer by a factor of seven.\cite{JiangJW2013bend} As a result, we find that the critical buckling strain for the MoS$_2$/MoS$_2$ is much larger than the graphene/graphene structure.

\begin{figure}[htpb]  
  \begin{center}
    \scalebox{1.0}[1.0]{\includegraphics[scale=1.0]{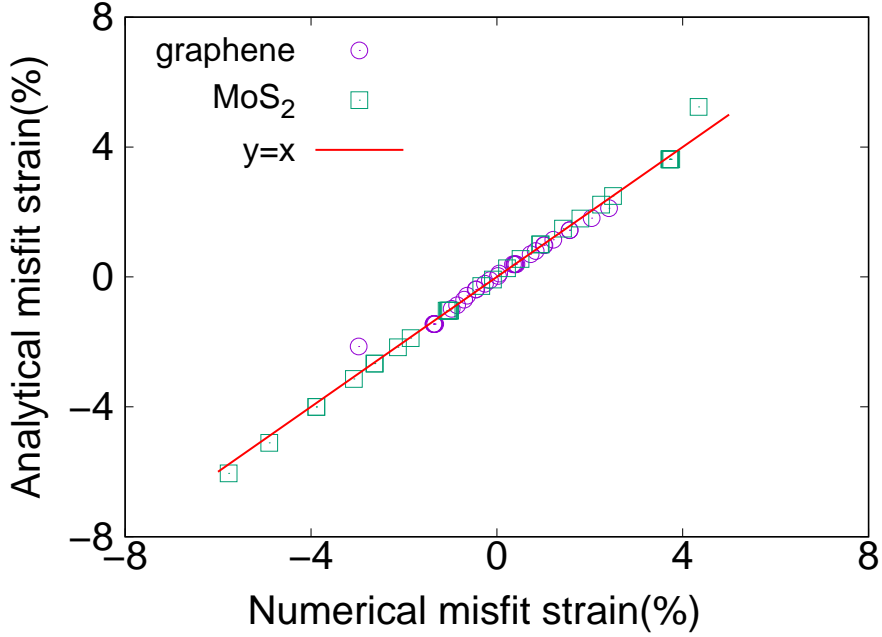}}
  \end{center}
  \caption{The misfit strain in the graphene/MoS$_2$ heterostructures.}
  \label{fig_misfit_strain}
\end{figure}

The length dependence for the buckling strain of these three types of structures are shown in Fig.~\ref{fig_lsi}. The buckling strain for the graphene/graphene and MoS$_2$/MoS$_2$ structures can be described by the Euler buckling theory. Some deviations from the Euler buckling theory are probably due to the high strain rate as discussed previously.\cite{Jiang2014The} For the graphene/MoS$_2$ heterostructure, an overall information from Fig.~\ref{fig_lsi} is that the critical buckling strain for the graphene/MoS$_2$ structure is between the values of MoS$_2$/MoS$_2$ and graphene/graphene, and are closely related to the misfit strain. To disclose the mechanism for the misfit strain effect on the buckling phenomenon, we first derive the analytic formula for the misfit strain in the graphene/MoS$_2$ heterostructure. The total strain energy of a heterostructure related to the misfit strain is\cite{alred2015interface}
\begin{eqnarray}
U =\frac{1}{2}E_1 \epsilon_1^2 + \frac{1}{2}E_2 \epsilon_2^2,
\label{eq_s1}
\end{eqnarray}
where $\epsilon_1=\frac{l-l_1}{l_1}$ and $\epsilon_2=\frac{l-l_2}{l_2}$ are the strain for these two atomic layers. Here $l$ is the final length of the heterostructure structure, while $l_1$ and $l_2$ are the original lengths for graphene and MoS$_2$ layer, respectively. The Young's moduli are $E_1$ and $E_2$. The final length $l$ can be obtained by minimizing the total strain energy with respective to $l$,
\begin{eqnarray}
l=\frac{E_1 l_2 + E_2 l_1}{E_1\frac{l_2}{l_1} + E_2\frac{l_1}{l_2}}.
\label{eq_s2}
\end{eqnarray}
The misfit strain is defined to be the strain in the graphene layer as follows
\begin{eqnarray}
\epsilon_m = \epsilon_1 = - \frac{l_1 - l}{l_1}.
\label{eq_misfit}
\end{eqnarray}
Fig.~\ref{fig_misfit_strain} shows that the misfit strain predicted by the analytic expression in Eq.~(\ref{eq_misfit}) agrees quite well with numerical simulation results. For comparison, the strain in the MoS$_2$ layer, $\epsilon_2 = - \frac{l_2 - l}{l_2}$, is also plotted in the figure.

\begin{figure}[htpb]  
  \begin{center}
    \scalebox{1.0}[1.0]{\includegraphics[scale=1.0]{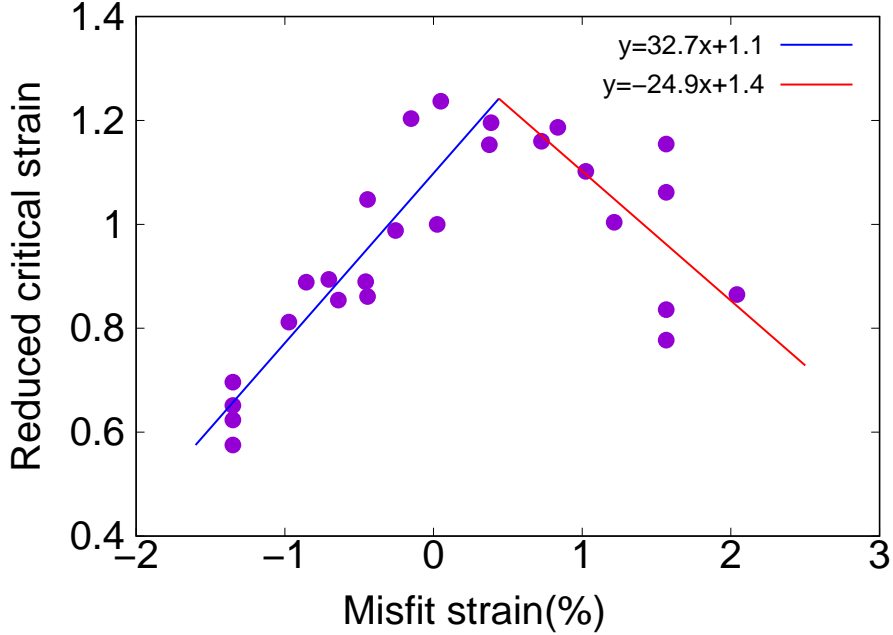}}
  \end{center}
  \caption{The relation between the reduced critical buckling strain and the misfit strain of the graphene/MoS$_2$ heterostructure. See text for the definition of the reduced critical buckling strain.}
  \label{fig_misfit}
\end{figure}

\begin{figure}[htpb]  
  \begin{center}
    \scalebox{1.0}[1.0]{\includegraphics[scale=0.4]{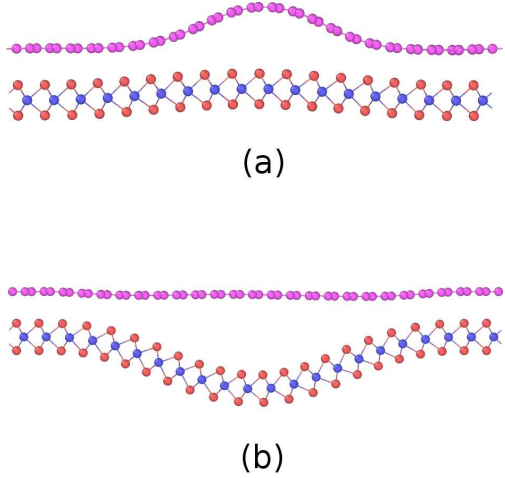}}
  \end{center}
  \caption{Buckling modes for the graphene/MoS$_2$ heterostructure with misfit strain of large magnitudes (a) -1.4\% and (b) 2.0\%.}
  \label{fig_debond}
\end{figure}

To demonstrate the misfit strain effect on the buckling phenomenon, we perform a set of simulations for the graphene/MoS$_2$ heterostructure with length varied within the range of [40, 160]~{\AA}. The misfit strain for these structures is within the range of [-2\%, 2\%]. Length and misfit strain both have effects on the critical buckling strain. To reveal the misfit strain effect separately, we get rid of the length effect by using the length $L$ to scale the critical strain, i.e., $\epsilon^L_c=\epsilon_c L^2$. The quantity $\epsilon^L_c$ should be length-independent according to the Euler buckling theory, so this quantity solely reflects the misfit strain effect. We further introduce a reduced critical strain $\tilde{\epsilon}_c=\epsilon^L_c/\epsilon^L_{c0}$, in which $\epsilon^L_{c0}$ is the scaled critical strain for the heterostructure structure with minimum (almost zero) misfit strain. With this definition, the critical buckling strain is reduced (upgraded) by the misfit strain if $\tilde{\epsilon}_c<1$ ($\tilde{\epsilon}_c>1$). There is no effect from the misfit strain for $\tilde{\epsilon}_c=1$.

Figure~\ref{fig_misfit} shows the relation between the reduced critical buckling strain and the misfit strain. The reduced critical buckling strain varies from 0.6 to 1.2 for different misfit strains; i.e., the misfit strain can affect the buckling strain by a factor of two. There are several interesting features in the figure. For negative misfit strain, $\epsilon_m\in[-2\%, 0]$, graphene is pre-compressed while MoS$_2$ is pre-stretched according to the definition of the misfit strain in Eq.~(\ref{eq_misfit}). The critical buckling strain decreases with the increase of the magnitude of the misfit strain. It should be noted that the buckling strain for graphene is much smaller than MoS$_2$. The buckling of the graphene/MoS$_2$ heterostructure thus initiates from the graphene layer. For negative misfit strain, the graphene layer in the graphene/MoS$_2$ heterostructure is pre-compressed by the misfit strain, so the graphene layer can be buckled even more easily, resulting in the reduction of the critical buckling strain. If the magnitude of the misfit strain is too large, then the graphene layer is seriously pre-compressed while the MoS$_2$ layer is considerably pre-stretched. As a result, the pre-compressed graphene layer is readily to be buckled, but the pre-stretched MoS$_2$ is difficult to be buckled. Eventually, the decoupling between these two atomic layers occurs as shown in Fig.~\ref{fig_debond}~(a).

If the misfit strain is in the small positive range, $\epsilon_m\in[0, 0.5\%]$, the graphene layer in the graphene/MoS$_2$ heterostructure is pre-stretched and the MoS$_2$ layer is pre-compressed by the misfit strain. The buckling of the heterostructure still initiates from the buckling of the graphene layer. However, the pre-stretched graphene layer becomes more difficult to be buckled, i.e., larger compressive strain is needed to cause the buckling of the graphene layer. As a result, the critical buckling strain increases with the increase of the magnitude of the misfit strain in the small positive region [0, 0.5\%].

For misfit strain in a large positive range, $\epsilon_m\in[0.5\%, 2\%]$, the graphene layer in the heterostructure is pre-stretched while the MoS$_2$ layer is pre-compressed by the misfit strain. The pre-compression in the MoS$_2$ layer is so large that the buckling of the graphene/MoS$_2$ heterostructure is initiated by the MoS$_2$ layer instead of the graphene layer. For larger misfit strain in this positive range, the MoS$_2$ layer is pre-compressed more seriously, so the buckling of the MoS$_2$ layer becomes easier. As a result, the critical buckling strain for the graphene/MoS$_2$ heterostructure decreases with increasing misfit strain in [0.5\%, 2\%]. For positive misfit strain of large magnitude, the decoupling can also happen as shown in Fig.~\ref{fig_debond}~(b). It is because the seriously pre-stretched graphene layer is difficult to be buckled, while the pre-compressed MoS$_2$ layer is easily to be buckled. The competition between these two atomic layers leads to the decoupling phenomenon.

The above effects can be discussed qualitatively in a more general scheme. For $\epsilon_m\in[-2\%, 0]$ and $\epsilon_m\in[0, 0.5\%]$, the buckling of the graphene/MoS$_2$ heterostructure initiates from the graphene layer, so the buckling strain is mainly governed by the graphene layer, i.e., $\tilde{\epsilon}_c \propto (\epsilon_c^{\rm gra} + \epsilon_m )$ with $\epsilon_c^{\rm gra}$ as the buckling strain for graphene. For $\epsilon_m\in[0.5\%, 2\%]$, the buckling of the graphene/MoS$_2$ heterostructure initiates from the MoS$_2$ layer, and the buckling strain is mainly governed by the MoS$_2$ layer, i.e., $\tilde{\epsilon}_c \propto (\epsilon_c^{\rm MoS_2} - \epsilon_m )$ with $\epsilon_c^{\rm MoS_2}$ as the buckling strain for MoS$_2$. Indeed, Fig.~\ref{fig_misfit} shows that the numerical data can be well fitted to two linear functions in different ranges.

\begin{figure}[htpb]  
  \begin{center}
    \scalebox{1.0}[1.0]{\includegraphics[scale=1]{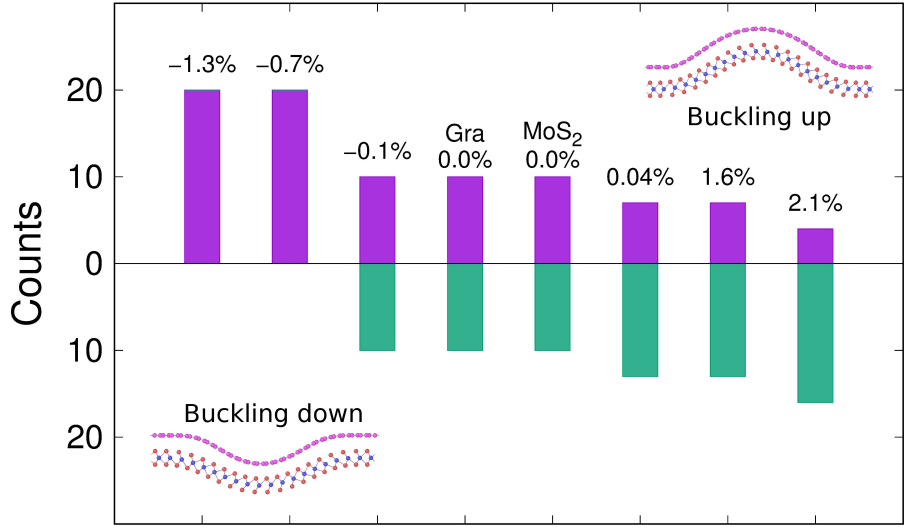}}
  \end{center}
  \caption{Probability for the buckling toward the graphene side (purple histograms) or the MoS$_2$ side (green histograms) in the graphene/MoS$_2$ heterostructure with different misfit strain. Top inset shows the buckling shape towards the upper graphene side. Bottom inset shows the buckling shape towards the lower MoS$_2$ side.}
  \label{fig_histogram}
\end{figure}

The misfit strain also affects the buckling direction for the graphene/MoS$_2$ heterostructure as shown in Fig.~\ref{fig_histogram}. More specifically, the heterostructure can buckle toward either the upper side or the lower side. This buckling direction depends on the misfit strain for the graphene/MoS$_2$ heterostructure. For each structure, we perform twenty MD simulations with different random distributions for the initial atomic velocities, while all other simulation parameters are the same. We count the number for the structure to buckle upwards or downwards. For graphene/graphene and MoS$_2$/MoS$_2$ structures, the buckling shape has equal probability to buckle upwards or downwards. However, for graphene/MoS$_2$ heterostructure of large negative misfit strain, the buckling shape prefers to buckle toward the graphene side, because the pre-compressed graphene layer is easily to be buckled while the pre-stretched MoS$_2$ layer is difficult to be buckled. In contrast, for large positive misfit strain, the buckling shape prefers to buckle toward the MoS$_2$ side. The probability for the buckling direction varies gradually with increasing misfit strain. Generally, the heterostructure prefers to buckle towards the atomic layer that is pre-compressed by the misfit strain.

\section{Conclusion}
To summarize, we have performed MD simulations to study the buckling phenomenon of graphene/MoS$_2$ van der Waals heterostructures under uniaxial compression. We find that the misfit strain can affect both the critical buckling strain and the buckling direction for the heterostructure. The misfit strain has different effects in three strain ranges. For negative misfit strain with graphene layer pre-compressed, the critical buckling strain can be considerably reduced by increasing the magnitude of the misfit strain. In small positive misfit strain region, the critical buckling strain is upgraded by increasing misfit strain. The critical buckling strain is reduced by the increase of the magnitude of the misfit strain for large positive misfit strains. The effects of the misfit strain on the buckling phenomenon are discussed based on the competition between the buckling of the graphene layer and the MoS$_2$ layer.

\textbf{Acknowledgment} The work is supported by the National Natural Science Foundation of China (Grant Nos. 11822206 and 12072182) and Innovation Program of the Shanghai Municipal Education Commission (Grant No. 2017-01-07-00-09-E00019).

\end{document}